\documentclass[twoside]{dis04}
\def\runauthor{M. Giunta}
\def\shorttitle{Recent QCD results from OPAL}
\begin{document}

\newcommand {\as} { \alpha_S }
\newcommand {\Z} { Z^0 }
\newcommand {\ee} {e$^+$e$^-$}
\newcommand {\ecm} { E_{c.m.} }
\newcommand {\mnch} {\langle n_{{\mathrm ch.}} \rangle }
\newcommand {\ptlu} { p_{\perp} }
\newcommand {\ycut} { y_{{\mathrm cut}} }
\newcommand {\ymin} { y_{{\mathrm min}} }
\newcommand {\dygap} { \Delta y_{{\mathrm gap}} }
\newcommand {\qtot} {Q_{{\mathrm leading}}}
\newcommand {\elead} {E_{{\mathrm leading}}}
\newcommand {\mlead} {M_{{\mathrm leading}}}
\newcommand {\qjet}  {\kappa_{{\mathrm jet}}}

\title{RECENT QCD RESULTS FROM OPAL
}

\author{Marina Giunta}

\address{Physics Department, University of California,\\
        Riverside, CA 92521, USA\\
E-mail: marina.giunta@cern.ch }

\maketitle

\abstracts{Some recent QCD results from the OPAL Collaboration are summarized. In particular: a test of color reconnection models and a search for glueballs using gluon jets with a rapidity gap; a study of unbiased gluon jet properties using a new technique called jet boost algorithm; a measurement of the strong coupling constant $\as$ from radiative events.
}

\section{Introduction} 
All the following analyses use multihadronic $\Z$ decay events observed with the OPAL detector~\mbox{\cite{bib-detector,bib-si}} at the LEP \ee collider at CERN.
In the first section, results on color reconnection and glueball searches published in a recent paper~\cite{op-pr379} are summarized; section two presents a study~\cite{bib-unbiased} of unbiased gluon jet properties and the last section is dedicated to a new measurement~\cite{bib-alphas} of $\as$ from radiative events.  
\section{Gluon jets with a rapidity gap}
\subsection{Test of color reconnection models}
Color reconnection (CR), i.e  a rearrangement of the color structure of an event, can be a source of rapidity gap events.
In the standard Monte Carlo models CR is not included, so the color flux tube is stretched from a quark to the corresponding antiquark without crossing. If CR is included in the Monte Carlo, we can have  more complex diagrams, in which string segments can either cross or appear as disconnected entities whose endpoints are gluons. In events with an isolated gluonic system a rapidity gap can form between the particles coming from the hadronization of the isolated segment - often the leading (highest rapidity) part of  a gluon jet - and the rest of the event. Thus rapidity gaps in gluon jets provide a sensitive means to search for color reconnection effects.

OPAL used this idea to test three CR models: one  implemented in the Ariadne Monte Carlo~\cite{bib-lonnblad}, one in Herwig~\cite{bib-herwig2} and a model  introduced by Rathsman~\cite{bib-rathsman} in the Pythia Monte Carlo. We refer to these as the Ariadne-CR, Herwig-CR, and Rathsman-CR models, respectively.

All the events are forced to three jets using  the Durham~\cite{bib-durham}  jet finder with a variable value for the resolution scale $y_{cut}$. The two quark ($\mathrm{q~ or ~\bar{q}}$) jets are identified using a b-tagging technique~\cite{b-tag} and the remaining jet is the gluon jet.

\begin{center}
\begin{figure}
\hspace*{2cm}
\includegraphics[width=8.cm]{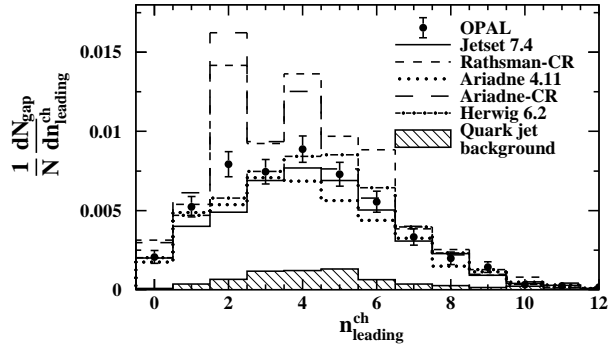}
\caption{Distributions of $n^{ch}_{leading}$  in the leading part of gluon jets. ``N'' represents the total number of selected gluon jets and ``$\mathrm{N_{gap}}$'' the number of gluon jets with a rapidity gap.} 
\label{fig_nch}
\end{figure}
\end{center}

The rapidity gap is defined using charged and neutral particles, requiring the smallest particle rapidity in a jet $y_{min}>1.4$, or, if this is not the case, the maximum difference between the rapidities of adjacent rapidity-ordered particles $\Delta y_{max}>1.3$ (this second sample is not used for Herwig-CR as explained in~\cite{op-pr379}).


The distributions of charged multiplicity ($n^{ch}_{leading}$) and total charge ($Q_{leading}$) of the leading part of gluon jets are normalized to the total number of selected jets before the rapidity gap requirement. The results for the $n^{ch}_{leading}$ distribution are shown in Fig.~\ref{fig_nch}. Both Rathsman-CR and Ariadne-CR predict a large excess of entries  at $n^{ch}_{leading}=2$ and 4 and also at $Q_{leading}=0$ (figure not shown). These are consequences of events with an isolated gluonic system in the leading part of the gluon jet, which is neutral and decays into an even number of charged particles. Herwig-CR predicts a less striking excess for $3\leq n^{ch}_{leading}\leq 5$ and $Q_{leading}=0$. 

A re-tuning of the Rathsman-CR and the Ariadne-CR models is also tried, but in both cases we cannot find a set of parameters allowing the models to describe the gluon jets distributions (i.e. $n^{ch}_{leading}$ and $Q_{leading}$) while continuing to provide a good description of the general features of inclusive $\Z$ decays. 
It is concluded that the Rathsman-CR and the Ariadne-CR models are both disfavored, while it is not possible to obtain a definite conclusion concerning the Herwig-CR model.

\subsection{Glueball search}
In~\cite{bib-ochs},
gluon jets with a rapidity gap are also proposed as
a potentially favorable environment
for the production of color singlet bound states of gluons,
such as glueballs.
Therefore, invariant mass spectra in the
leading part of the selected gluon jets are examined.
Since it is a  search for anomalous resonant structure,
the data are compared to the predictions of
the models without color reconnection,
i.e. Jetset, Ariadne and Herwig.
These models do not contain glueballs.
Three different distributions are examined: the total invariant mass of the leading part of the gluon jets with $Q_{leading}=0$ (since glueballs are electrically neutral), the distribution of invariant mass of two oppositely charged particles in the leading part of the gluon jets and the corresponding distribution of four charged particles with total electric charged zero.
No evidence for anomalous production of scalar particles was observed.

\section{Unbiased gluon jets' properties using the jet boost algorithm}
The Lund theory group proposed a new method, called \textit{jet boost algorithm}~\cite{bib-boost} (BA), to determine properties of unbiased gluon jets. 
After testing the method, OPAL used it to study unbiased gluon jets at different energies and then compared the results with theoretical predictions.

The BA is based on the color dipole model of QCD and can be summarized in the following way: in three-jet events, symmetric with respect to the gluon direction, the two independent dipoles (one connecting the $q$ and the $g$ and the other going from the $\bar{q}$ and the $g$) can be boosted and combined to yield the dipole structure of a $gg$ event.

Experimentally, three jets are reconstructed in each multihadronic event and the gluon jet is identified with the same technique used in the previous analysis, obtaining a  final sample of 25396 events with a gluon jet purity of 85\%. Then each event was boosted to a frame where the angle between the $q$ and the $g$ was the same as the angle between the $\bar{q}$ and the $g$ ($\theta_{qg}=\theta_{\bar{q}g}=\theta$), yielding a symmetric event. The multiplicity of the gluon jet is given by the number of particles lying inside the cone-like region defined by the bisectors of $\theta_{qg}$ and $\theta_{\bar{q}g}$ and the corresponding energy scale of the unbiased gluon jet, $E^*_g$, is equal to the transverse momentum of the gluon jet, defined by~\cite{bib-plund}:
\begin{equation}
p_{\perp, gluon}=\frac{1}{2} \sqrt{\frac{s_{qg}s_{\bar{q}g}}{s}}
\end{equation}
where $s_{ij}$ ($i,j=q,\bar{q},g$) is the invariant mass squared of the $ij$ pair and $s=E^2_{c.m.}$.   
The events are divided in 7 subsamples according to the gluon jet energy $E^*_g$.

Using Herwig 6.2, we compare results obtained with the BA method on simulated three-jet events with results for unbiased gluon jets from simulated color singlet $gg$ events. We compare the multiplicity distributions (in the 7 energy intervals) and we found a good agreement for $E^*_g>5$ GeV; we also measured the fragmentation functions which agreed for $E^*_g>14$ GeV. 

\begin{center}
\begin{figure}
\hspace*{2cm}
\includegraphics[width=8.cm]{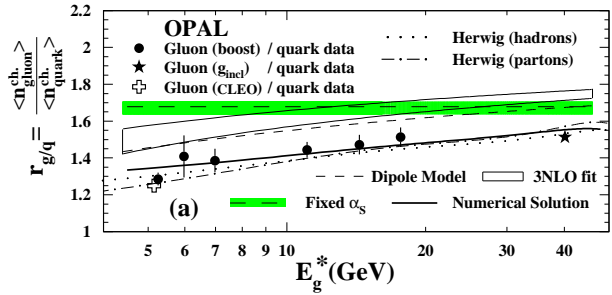}
\caption{The ratio between the mean charged particle multiplicities of unbiased gluon and uds flavored quark jets, $r_{q/g}$, as a function of jet energy $E^*_g$.} 
\label{fig-ratio}
\end{figure}
\end{center}

After establishing the validity of the method, many measurements of unbiased gluon jet properties are performed: the mean multiplicity (in the 7 energy bins), actually the most precise result for $5.25<E^*_g<20$ GeV; the first measurement  of the factorial moments $F_2$ and $F_3$ of unbiased gluon jets multiplicity distributions over an energy range; the ratio between the mean charged particle multiplicities of unbiased gluon and uds flavored quark jets, $r_{q/g}$, as a function of jet energy (see Fig.~\ref{fig-ratio}). The measurements are compared to many different theoretical predictions and an overall global agreement was found. 
Finally, the fragmentation function was measured for jets with $E^*_g>14$ GeV and was fitted using the DGLAP evolution equation. The fit gives a good description of the measurements and yields a result for the strong coupling constant consistent with the world average:
\begin{equation}
\alpha_s(m_Z) = 0.128\pm 0.008(stat.) \pm 0.015(syst.)
\end{equation}
 which provides a unique consistency check of QCD.   

\section{Measurement of $\as$ from radiative events}
Hadronic final states with a high energy isolated photon (\ee $\rightarrow q\bar{q}\gamma$) are selected and the strong coupling constant $\as$ is extracted by fitting event shape variables for the reduced center-of-mass energies ranging from 20 to 80 GeV. This procedure is allowed by assuming that photons emitted before or immediately after the $\Z / \gamma$ production do not interfere with QCD processes. 

Our signal is given by multihadronic events with ISR/FSR high energy photons ($E_{\gamma} > 10$ GeV). The possible backgrounds are: non-radiative multihadronic (NRMH)  events, two photon processes and $\tau$ pair production.

Some isolation conditions are imposed on electromagnetic clusters in order to have well isolated high energy photons: the cluster must have an energy larger than 10 GeV, be located in the barrel region (i.e. $|\cos(\theta_{EC})|<0.72$) of the detector; the angle with respect to the axis of any jet, $\alpha^{iso}_{jet}$, is required to be larger than 25$^\circ$,  the sum of tracks momenta and the sum of energy deposition in the electromagnetic calorimeter within a cone of 0.2 radiant around the candidate must be both less than 0.5 GeV.
After the isolation cuts 11625 clusters are retained, but the background from NRMH events is still large, 52.8\%, due to clusters arising from $\pi^0$ decay.

In order to further reduce this background a likelihood ratio method with four input variables ($|\cos(\theta_{EC})|$, $\alpha^{iso}_{jet}$, cluster shape fit variable, distance between electromagnetic cluster and associated presampler cluster) is applied. The events are divided in seven subsamples according to the cluster energy ($E_{cluster}$) and the cut on the likelihood value is chosen in order to keep 80\% of signal events.

As mentioned in~\cite{bib-background}, JETSET fails to reproduce the observed rate of isolated electromagnetic clusters, so the background fraction was estimated directly from our data by two independent methods: by fitting the likelihood distributions in the data with a linear combination of the Monte Carlo distributions for signal and background events or by assuming isospin symmetry (charged hadrons satisfying the same isolation cut criteria are selected and the rate of isolated neutral hadrons are obtained from the rate of isolated charged hadrons).
The final background due to NRMH events is $<10$\% for $E_{cluster}$=10-35 GeV and 10-15\% for $E_{cluster}$=35-45 GeV, the two photon processes are $<0.01$\% and the $\tau\tau$ events are 0.5-1.0\%.

After boosting the hadronic system into the center-of-mass frame of the hadrons, the following event shape variables are calculated: Thrust ($T$), Heavy Jet Mass ($M_H$) and Jet Broadening Variables ($B_T$ and $B_W$). The normalized background distributions are subtracted from data at detector level and then, in order to correct for detector effects (acceptance, resolution), bin-by-bin correction factors are applied and distributions at the so called hadron level are obtained.
The determination of $\as$ is obtained by fitting perturbative QCD predictions to the event shape distributions corrected to the hadron level. The $O(\as^2)$ and NLLA calculations are combined with the $\ln(R)$ matching scheme and corrected to hadron level multiplying by a hadronization correction factor obtained using Monte Carlo informations.
The fit to the event shape variables uses a least $\chi^2$ method with $\as(Q)$ treated as a free parameter. The regions with small and uniform corrections are selected as fitting ranges. 
Values of $\as$ for each event shape variable and for all event shape variables combined are fitted to the solution of the renormalization group equation at NNLO described in~\cite{bib-rge}. The following value of $\Lambda^{(5)}_{\bar{MS}}$ is derived from the fit:
\begin{equation}
\Lambda^{(5)}_{\bar{MS}}=0.2027\pm0.0141(stst.)^{+0.1130}_{-0.0939}(syst.)~\mathrm{GeV}~(\chi^2/d.o.f.=33.1/6)
\end{equation}
All values of $\as$ are then propagated to the energy scale of $M_Z$ and a combined value of $\as$ for all the reduced center-of-mass energy subsamples and  all event shapes variables is extracted:  
\begin{equation}
\as(M_Z)=0.1176\pm0.0012(stat.)^{+0.0093}_{-0.0085}(syst.)
\end{equation}
This agrees with the previous OPAL analysis using non-radiative LEP1 (i.e. $E_{c.m.}=91$ GeV) data and the PDG world average.


\begin{thebibliography}{0}
\bibitem{bib-detector}
   OPAL, K. Ahmet {\it et al.},
   Nucl. Instr. and Meth. {\bf A305} (1991)~275.
\bibitem{bib-si}
   P.P. Allport {\it et al.},
   Nucl. Instr. and Meth. {\bf A346} (1994)~476.
\bibitem{op-pr379} OPAL, G. Abbiendi {\it et al.}, accepted by Eur. Phys. J. C, hep-ex/0306021. 
\bibitem{bib-unbiased} OPAL, G. Abbiendi {\it et al.}, Phys. Rev. {\bf D69} (2004) 032002.
\bibitem{bib-alphas} OPAL Physics Note OPAL-PN519 (April 2003).
\bibitem{bib-lonnblad}
   L. L\"{o}nnblad, Z. Phys. {\bf C70} (1996)~107.
\bibitem{bib-herwig2}
   G. Corcella {\it et al.}, JHEP 0101 (2001)~010.
\bibitem{bib-rathsman}
   J. Rathsman, Phys. Lett. {\bf B452} (1999)~364.
\bibitem{bib-durham}
   S. Catani et al., Phys. Lett. {\bf B269} (1991)~432.
\bibitem{b-tag}OPAL, R. Akers  {\it et al}, Z. Phys.{\bf C68} (1995)~179.
\bibitem{bib-ochs}
   P. Minkowski and W. Ochs, Phys. Lett. B485 (2000)~139. 
\bibitem{bib-boost} P. Eden, G. Gustafson, JHEP {\bf 9809} (1998) 015.
\bibitem{bib-plund} P. Eden, G. Gustafson and V.A. Khoze, Eur. Phys. J. C 11 (1999) 345.
\bibitem{bib-background} OPAL, K. Ackerstaff {\it et al}, Eur. Phys. J. C 5 (1998) 411; \\L3, O. Adriani {\it et al},  Phys. Lett. {\bf B292} (1992)~472.
\bibitem{bib-rge} Particle Data Group, K. Hagiwara {\it et al},  Phys. Rev. {\bf D66} (2002) 010001.
\end{thebibliography}
\end{document}